\pdfoutput=1
\PassOptionsToPackage{bookmarks=false,colorlinks=true,linkcolor=blue,urlcolor=cyan,filecolor=magenta,citecolor=red,pdfstartview=FitV,hyperfootnotes=false,pdftitle={Tantum gravity},pdfauthor={Florian Ecker, Adrien Fiorucci, Daniel Grumiller},pdfsubject={Tantum gravity},pdfkeywords={Tantum gravity},bookmarksopen=true}{hyperref}
\documentclass[10pt,twocolumn,superscriptaddress,aps,prl]{revtex4}

\usepackage{amsmath,amsfonts,amssymb,amsthm,amstext,amscd,eucal,graphicx,xcolor,orcidlink,ulem}
\usepackage[utf8]{inputenc}
\usepackage[english]{babel}
\usepackage{graphicx}
\usepackage{dcolumn}
\usepackage{bm}
\usepackage{braket}
\usepackage{mathrsfs}
\usepackage{tikz}
\usetikzlibrary{decorations.pathmorphing,backgrounds,calc,shapes,arrows,shapes.misc,shadows,shadings}
\tikzset{cross/.style={cross out, draw=black, minimum size=2*(#1-\pgflinewidth), inner sep=0pt, outer sep=0pt},
cross/.default={1pt}}

\definecolor{darkgreen}{rgb}{0,0.3,0}
\definecolor{darkblue}{rgb}{0,0,0.3}
\definecolor{darkred}{rgb}{0.7,0,0}

\DeclareMathOperator{\extdm}{d}
\newcommand{\dd}{\extdm \!}

\begin{document}


\title{Tantum Gravity}

\author{\orcidlink{0000-0002-0449-0081}Florian Ecker}
\email{fecker@hep.itp.tuwien.ac.at}
\affiliation{Institute for Theoretical Physics, TU Wien, Wiedner Hauptstrasse 8–10/136, A-1040 Vienna,
Austria}

\author{\orcidlink{0000-0002-4520-6286}Adrien Fiorucci}
\email{adrien.fiorucci@tuwien.ac.at}
\affiliation{Institute for Theoretical Physics, TU Wien, Wiedner Hauptstrasse 8–10/136, A-1040 Vienna,
Austria}

\author{\orcidlink{0000-0001-7980-5394}Daniel Grumiller}
\email{grumil@hep.itp.tuwien.ac.at}
\affiliation{Institute for Theoretical Physics, TU Wien, Wiedner Hauptstrasse 8–10/136, A-1040 Vienna,
Austria}

\date{\today}

\begin{abstract} 
We argue there is an interesting triple-scaling limit of quantum gravity, namely when Planck's constant scales to infinity while Newton's constant and the speed of light tend to zero, keeping fixed the gravitational coupling $G_N\,c^{-4}$ and the combination $\hbar\,c$. We refer to this limiting theory as ``tantum gravity'' and describe in this Letter some of its main properties and prospects for physics. Most notably, the laws of black hole thermodynamics survive this limit, which means that puzzles related to black holes and their evaporation could be addressed more easily in tantum gravity than in fully-fledged quantum gravity.
\end{abstract}

\maketitle






The word ``quantum'' is so ingrained in our language that it can be easy to forget its etymology: the Latin ``quantum'' means ``how much?'' --- a fitting label for the elusive gravity theory we collectively try to grapple with. So far, quantum gravity is best understood in scaling limits where either Newton's constant vanishes (quantum field theory) or Planck's constant vanishes (Einstein gravity). As a further simplification, a large speed of light is often assumed (Galilean limit), which suppresses particle creation in quantum field theories and reduces Einstein gravity to Newton gravity. 

\begin{figure}[hbt]
\def\L{3.5}
\begin{center}
\begin{tikzpicture}[scale=0.8]
\draw[black,thin,->] (0,0) coordinate (orig) -- (0,1.1*\L);
\draw[black,thin,->] (orig) -- (1.1*\L,0);
\draw[black,thin,->] (orig) -- (-0.55*\L,-0.55*\L);
\draw[red,thin,-] (orig) -- (0,\L) coordinate (y);  
\draw[blue,thin,-] (orig) -- (\L,0) coordinate (x);
\draw[green,thin,-] (orig) -- (-0.5*\L,-0.5*\L) coordinate (z);
\draw[purple,thin,-] (y) -- (\L,\L) coordinate (xy);
\draw[violet,thin,-] (x) -- (xy);
\draw[cyan,thin,-] (z) -- (0.5*\L,-0.5*\L) coordinate (xz);
\draw[teal,thin,-] (x) -- (xz);
\draw[lime,thin,-] (z) -- (-0.5*\L,0.5*\L) coordinate (yz);
\draw[pink,thin,-] (y) -- (yz);
\draw[yellow,thin,-] (yz) -- (0.5*\L,0.5*\L) coordinate (xyz);
\draw[olive,thin,-] (xz) -- (xyz);
\draw[magenta,thin,-] (xy) -- (xyz);
\draw[darkgray,thick] (xz) circle(0.02*\L);
\draw[black] (xz) node[right,yshift=-0.5em] {TG};
\draw[darkgray,thick] (y) circle(0.02*\L);
\draw[black] (y) node[right,yshift=-0.5em] {TG$^\ast$};
\draw[red] (-0.25*\L,0.25*\L) node[rotate=45] {GR};
\draw[blue] (0.25*\L,-0.25*\L) node[rotate=22.5] {QFT};
\draw[black] (x) node[right,xshift=1.0em] {$\hbar$};
\draw[black] (y) node[above,yshift=1.0em] {$G_N$};
\draw[black] (z) node[left,yshift=-1.0em] {$c^{-1}$};
\draw[black] (orig) node[above,xshift=0.5em] {0};
\draw[black] (orig) node[below,xshift=0.2em] {0};
\draw[black] (orig) node[left,yshift=0.3em] {0};
\draw[black] (x) node[above,xshift=-0.6em] {$\infty$};
\draw[black] (y) node[left,xshift=0em] {$\infty$};
\draw[black] (z) node[left,yshift=0.3em] {$\infty$};
\end{tikzpicture}
\end{center}
\vspace*{-0.6truecm}
\caption{Bronstein cube with tantum gravity limit highlighted as TG and its antipodal as TG$^\ast$.}
\label{fig:1}
\end{figure}
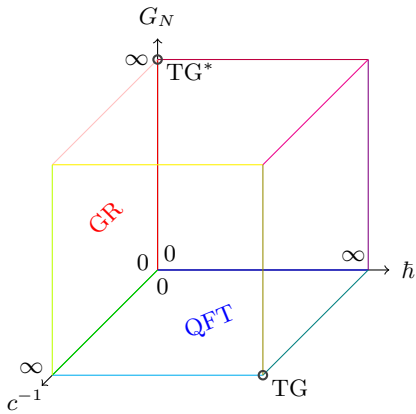

These various limits can be captured by the Bronstein cube, depicted in Fig.~\ref{fig:1}. The three axes in the Bronstein cube are Newton's constant $G_N$, Planck's constant $\hbar$, and the inverse vacuum speed of light $c^{-1}$. Unlike traditional renderings, we have compactified all axes so that the faces, edges, and corners of the cube correspond to zero or infinite values of the corresponding coupling constants. Quantum gravity fills the volume of the cube, and each of the limiting cases corresponds to a face, a double-scaling limit to an edge, and a triple-scaling limit to a corner. This means that we have (at least) $6+12+8=26$ limiting theories of quantum gravity: the latter are listed in Table \ref{tab:1}, where the entries $0$, $1$, or $\infty$ mean that the corresponding quantity is zero, finite, or infinite.

The first nine limiting theories have common acronyms, \textit{i.e.}~GM stands for Galilean mechanics, SR for special relativity, QM for quantum mechanics, NG for Newtonian gravity, QFT for quantum field theory, GR for general relativity, GQG for Galilean quantum gravity, CM for Carrollian mechanics and CQM for Carrollian quantum mechanics. The limits 10-24 have no common names yet. The final two entries are tantum gravity (denoted as \textbf{TG}) and its dual limit (denoted as TG$^\ast$).

\begin{table}[hbt]
\begin{tabular}{c|ccccccccc}
& GM & SR & QM & NG & QFT & GR & GQG & CM & CQM \\\hline
$G_N$ & 0 & 0 & 0 & 1 & 0 & 1 & 1 & 0 & 0 \\
$\hbar$ & 0 & 0 & 1 & 0 & 1 & 0 & 1 & 0 & 1  \\
$c^{-1}$ & 0 & 1 & 0 & 0 & 1 & 1 & 0 & $\infty$ & $\infty$ \\\hline\hline
& 10 & 11 & 12 & 13 & 14 & 15 & 16 & 17 & 18 \\\hline
$G_N$ & $\infty$ & $\infty$ & $\infty$ & 1 & $\infty$ & 1 & 1 & $\infty$ & $\infty$ \\
$\hbar$ & $\infty$ & $\infty$ & 1 & $\infty$ & 1 & $\infty$ & 1 & $\infty$ & 1 \\
$c^{-1}$ & $\infty$ & 1 & $\infty$ & $\infty$ & 1 & 1 & $\infty$ & 0 & 0 \\\hline\hline
& 19 & 20 & 21 & 22 & 23 & 24 & TG$^\ast$ & \textbf{TG} & \\\hline
$G_N$ & 1 & 1 & $\infty$ & 0 &
0 & $\infty$ & $\infty$ & $\boldsymbol{0}$ & \\
$\hbar$ & $\infty$ & 0 & 0 & $\infty$ & $\infty$ & 0 & 0 & $\boldsymbol{\infty}$ & \\
$c^{-1}$ & 0 & $\infty$ & 1 & 1 & 
0 & $\infty$ & 0 & $\boldsymbol{\infty}$ & \\
\end{tabular}
\caption{26 limits of quantum gravity.}
\label{tab:1}
\end{table}

It can be rewarding to go through the 26 entries in Table \ref{tab:1} and specify how the limits are taken, which combinations of $G_N$, $\hbar$, and $c$ remain finite, which sector of quantum gravity this theory maintains, which physical effects can be described by the limiting theory, \textit{etc}. In this manner, one may find more than one way of taking such limits for the same entry in the Table, so there could be more than 26 limiting theories, though not of equal interest for applications. Extending the Bronstein cube by additional axes (cosmological constant $\Lambda$, Boltzmann constant $k_{\textrm{\tiny B}}$ \cite{Cohen-Tannoudji:2009hwn}, number of degrees of freedom $N$ \cite{Oriti:2018tym}, \textit{etc}.) can also be fruitful but is unnecessary for our purposes.

The focus of this Letter is the highlighted entry in Table \ref{tab:1} or, equivalently, the corner in Fig.~\ref{fig:1} labeled ``TG,'' corresponding to infinite Planck's constant and vanishing Newton's constant and speed of light. We call this theory ``tantum gravity,'' where the Latin word ``tantum'' means ``that much!'' --- a fitting answer to the question ``quantum?''. 

However, it is not obvious that such a limit leads to a gravity theory. After all, we send Newton's constant to zero. To investigate this issue, we need to be more specific about these limits. We employ an inductive approach.

Our main motivation comes from black hole thermodynamics, which lies at the heart of many quantum gravity mysteries, including the information paradox, black hole microstates, and the holographic principle (see, \textit{e.g.}, \cite{'tHooft:1993gx,Susskind:1995vu,Maldacena:1997re,Gubser:1998bc,Witten:1998qj,Ryu:2006bv,Mathur:2009hf,Raju:2020smc,Almheiri:2020cfm} and references therein). Thus, we check under which conditions the Schwarzschild radius, entropy, temperature, and energy remain finite. 

Let us start with the latter. For a given black hole of mass $M$, its energy $E$ is given by $E=Mc^2$. Thus, we leave $c$ and $M$ finite or scale $M$ inversely to $c^2$. Written in terms of energy, the formula on Hawking's tombstone (with Boltzmann's constant set to one) 
\begin{equation}
    T = \frac{\hbar\,c^5}{8\pi\,G_N\,E} 
    \label{eq:tg10}
\end{equation}
shows that any single scaling limit in the Bronstein cube leads to infinite or zero Hawking temperature. Since we want to avoid this scenario, we deduce that our desired theory can only live at the edges or corners of the Bronstein cube and not at one of its faces. This eliminates, in particular, GR and QFT. 

At this stage, we still have many possibilities for double- and triple-scaling limits that maintain finite Hawking temperature (for instance, the Newton gravity limit $\hbar\to 0$, $c\to\infty$). Finiteness of the Bekenstein--Hawking entropy does not add any new conditions due to the Smarr formula $E=2TS$ or the first law $\delta E = T\,\delta S$, see for instance \cite{Bardeen:1973gs}.

Finiteness of the Schwarzschild radius $r_{\textrm{\tiny S}}=2G_N\,E/c^4$ imposes a second constraint, namely finiteness of $G_N/c^4$, implying that $\hbar\,c$ needs to remain finite to maintain finite Hawking temperature \eqref{eq:tg10}. This condition removes all the edges from the Bronstein cube and leaves us with the two marked points in Fig.~\ref{fig:1}, one of which is the tantum gravity limit $G_N,c\to 0$, $\hbar\to\infty$ and the other its dual version where all limits are taken oppositely. In the Bronstein cube, the latter corresponds to the antipodal point labeled as TG$^\ast$ in Fig.~\ref{fig:1}~\footnote{%
We keep finite the Planck length and energy in both limits but send the Planck time and mass to $\infty$ in tantum gravity and to $0$ in TG$^{\ast}$.
}.

We stress that the two points TG and TG$^\ast$ are the only points on the whole Bronstein cube (besides its quantum gravity interior) where all three thermodynamic quantities and the Schwarzschild radius are finite. So from the perspective of black hole thermodynamics, the tantum gravity limit is unique, up to dualization.

In summary, to obtain tantum gravity from quantum gravity, we have to take the Carrollian limit $c\to 0$ and retain finite combinations
\begin{equation}
    G_M := G_N\, c^{-4}\qquad\qquad \kappa :=   \hbar\,c~ .
    \label{eq:tg1}
\end{equation}
The first one is the gravitational coupling constant in front of the Einstein--Hilbert action. So it is not automatically true that sending $G_N\to 0$ removes gravity --- it is still there if simultaneously the speed of light is sent to zero while keeping $G_M$ fixed, which is precisely what is done to reach the so-called magnetic Carroll gravity limit \cite{Hansen:2021fxi,Campoleoni:2022ebj}, hence the notation $G_M$. This elementary observation justifies calling our limit theory ``gravity.'' Similarly, the limit $\hbar\to\infty$ does not automatically imply a breakdown of the semi-classical approximation since we still have two combinations of the coupling constants \eqref{eq:tg1} that could be either large or small. We shall demonstrate this explicitly below. Thus, one should consider the points in the Bronstein cube labeled as TG and TG$^\ast$ as two-dimensional planes spanned by the coupling constants $G_M$ and $\kappa$.

An efficient way to parametrize the triple scaling limit leading to TG is to rescale 
\begin{equation}
c\to c\,\epsilon\qquad\qquad\hbar\to\hbar/\epsilon\qquad\qquad G_N\to G_N\,\epsilon^4
\label{eq:whynot}
\end{equation}
and then take the limit $\epsilon\to0$. This has the added advantage that the limiting parameter is dimensionless, and we shall use this procedure below. The parameter $\epsilon$ drops out in the TG coupling constants \eqref{eq:tg1}. 

By construction, the tantum gravity limit not only leads to a gravity theory, but the limiting theory also contains black-hole-like states with finite entropy, temperature, and energy, related to each other by the first law. At first glance, this is again surprising since we take a Carrollian limit, $c\to 0$, for which the lightcone collapses \cite{Levy1965, SenGupta1966OnAA} and makes obsolete all notions of horizons. However, as we have seen above, one has to be careful with naive arguments based on singular limits. Indeed, as explained in \cite{Ecker:2023uwm}, there can be ``Carroll black holes,'' though one needs to extend the notion of black holes and avoid basing them on event horizons. Instead, they are defined by their thermal properties and the existence of a so-called Carroll extremal surface, which is reminiscent of the bifurcation sphere of the Schwarzschild black hole \cite{Ecker:2023uwm}. 

From a quantum gravity perspective, it is an asset that we are forced to go beyond the classical notion of a black hole in tantum gravity, since quantum black holes should not be defined in terms of event horizons either --- after all, quantum black holes evaporate \cite{Hawking:1974rv,Hawking:1975vcx}, so event horizons are artifacts of the classical approximation. This is one of several ways in which tantum gravity is closer to quantum gravity than classical gravity.

In the remainder of this Letter, we substantiate our assertions through a straightforward yet enlightening calculation. We delve into a specific illustrative example based on the Euclidean path integral approach to Einstein gravity, as formulated by Gibbons and Hawking in \cite{Gibbons:1976ue}. Their pivotal insight involves expanding the complete gravitational action $\Gamma[g]$, including relevant boundary terms, around a classical saddle-point $g_{\textrm{\tiny cl}}$. Which boundary terms are relevant is determined by demanding a well-defined variational principle, $\delta\Gamma[g_{\textrm{\tiny cl}};\,\delta g]=0$, as well as a finite on-shell action $\Gamma[g_{\textrm{\tiny cl}}]$. Under these hypotheses, the Euclidean partition function
 $Z=\int \mathscr{D}g\, \exp (-\frac{1}{\hbar }\,\Gamma [g])$
is well-approximated classically by the exponential of the on-shell action,
$Z\approx \exp (-\frac{1}{\hbar }\,\Gamma[g_{\textrm{\tiny cl}}])$
provided $\hbar$ is sufficiently small. Moreover, we need to specify boundary conditions before reducing the partition function to its classical approximation. In the simplest case (on which we shall focus below), this amounts to fixing the periodicity of Euclidean time, which in turn can be interpreted as the inverse temperature \cite{Gibbons:1976ue}.

For Einstein gravity, the exponent in the Euclidean partition function is given by 
\begin{equation}
    -\frac{1}{\hbar}\,\Gamma =\frac{1}{\hbar }\,\frac{c^3}{16\pi G_N}\int _{\mathcal{M}_4}\dd^4x\sqrt{g^{(4)}}\,R^{(4)}+\frac{1}{\hbar}\,I_{\partial \mathcal{M}_4}
    \label{eq:tg15}
\end{equation}
where $I_{\partial\mathcal{M}_4}$ is a boundary action that guarantees a well-defined variational principle and a finite on-shell action. For simplicity, we assume spherical symmetry and prepare the metric suitably for a TG limit,
\begin{equation}
    \dd s^2_{(4)}=\big(\epsilon^2\tau _\mu \tau _\nu +e_\mu e_\nu\big)\,\dd x^\mu \dd x^\nu +\frac{4}{\lambda^2}\,X\,\dd \Omega^2_{S^2}\, ,
    \label{eq:tg16}
\end{equation}
with the limiting parameter $\epsilon$, the temporal einbein $\tau _\mu$, the spatial einbein $e_\mu$, a positive constant $\lambda$ of inverse length dimension, the dimensionless surface area/dilaton field $X$, and the usual line element of the round 2-sphere $\dd\Omega^2_{S^2}$. The indices $\mu,\nu$ range only over 0 and 1, corresponding to (Euclidean) time $x^0$ and radius $r:=x^1$.

It is a classic result that inserting the spherically symmetric ansatz \eqref{eq:tg16} into the action \eqref{eq:tg15} yields a specific two-dimensional (2d) dilaton gravity model (see \cite{Berger:1972pg} for the Hamiltonian formulation, \cite{Thomi:1984na} for the Lagrangian formulation, and \cite{Grumiller:2007ju} for a derivation of the boundary term, along the lines of \cite{Skenderis:2002wp})~\footnote{
Since, in general, dimensional reduction and quantization do not commute \cite{Frolov:1999an} (``dimensional reduction anomaly'') one should think of the spherically reduced theory as a separate quantum gravity model, the classical limit of which coincides with the classical limit of spherically symmetric Einstein gravity. In our case, the reduced theory is a 2d dilaton gravity model. For the TG limit and the leading order saddle-point contribution to the partition function the dimensional reduction anomaly plays no role since the latter is a 1-loop effect and a consequence of the UV behavior of the theory when coupled to matter.
} 
\begin{equation}
    \begin{split}
        -\frac{1}{\hbar}\,\Gamma &= \frac{1}{\hbar} \frac{c^3}{G_N\lambda ^2}\int _{\mathcal{M}}\dd^2 x\det(\epsilon\tau,e) \, \mathcal L_{\textrm{\tiny DG}} +\frac{1}{\hbar }\,I_{\partial\mathcal{M}} \, ,\\
        \mathcal L_{\textrm{\tiny DG}} &:=    XR+\frac{1}{2X} (\partial X)^2 +\frac{\lambda ^2}{2}\, , 
    \end{split}
    \label{eq:tg17}
\end{equation}
where $\mathcal{M}$ is a 2d manifold and $R$ is the 2d Ricci scalar. To evaluate the action \eqref{eq:tg17}, it is useful to introduce a cutoff $r=r_c$ on the radial coordinate, thereby introducing a boundary, the cutoff surface $\partial\mathcal{M}$, and then send the cutoff to infinity at the end of the calculation. This ensures that all intermediate results are finite. The model \eqref{eq:tg17} has as most general solution static metrics and dilatons parametrized by a single constant \cite{Grumiller:2002nm}: in Schwarzschild gauge $X(r) =(\frac{\lambda r}{2})^2$ and 
\begin{equation}
    \dd s^2=\epsilon^2\xi(r)\,(\dd x^0)^2+\frac{\dd r^2}{\xi (r)}\qquad \xi(r) =1-\frac{r_{\textrm{\tiny S}}}{r}\,.
    \label{eq:tg18}
\end{equation}
The asymptotic boundary is a dilaton isosurface with $X(r_c)\gg1$ ($r_c\gg r_{\textrm{\tiny S}}$) \footnote{%
Inserting the 2d solution \eqref{eq:tg18} back into the 4d metric \eqref{eq:tg16} recovers precisely the Schwarzschild solution.
}.

Now, we set up a canonical ensemble by fixing the proper length $\ell$ of the Euclidean cycle at the boundary,
\begin{equation}
    \lim_{r_c\to \infty}\epsilon \oint_{\partial \mathcal{M}} \!\!\!\dd x^\mu \,\tau_\mu  
   \overset{!}{=}\ell = \beta \,\hbar c =\beta\,\kappa 
    \label{eq:tg19}
\end{equation}
with the boundary volume form $\epsilon \tau $ induced by the choice of normal vector $n=\sqrt{\xi}\,\partial_r$ and $\ell=\beta \kappa$ on dimensional grounds. As usual in Euclidean field theories, $\beta =1/T$ and in the saddle-point approximation, the on-shell action for a given classical solution is related to its free energy $F$ by $\frac{1}{\hbar}\Gamma=\beta F$.

Since eventually we intend to insert on-shell configurations into the action \eqref{eq:tg17}, we express it already in terms of the coordinates used for the solutions \eqref{eq:tg18}, 
\begin{equation}
    -\frac{1}{\hbar}\,\Gamma = \frac{1}{G_M\kappa\lambda^2}\oint\limits_0^{\ell} \dd x^0 \int\limits_{r_{\textrm{\tiny S}}}^{r_c} \dd r \det(\tau,e)\,\mathcal L_{\textrm{\tiny DG}} + \frac{1}{\hbar}\, I_{\partial\mathcal M}
    \label{eq:tg20}
\end{equation}
with the boundary term \cite{Grumiller:2007ju}
\begin{equation}
        \frac{1}{\hbar}\,I_{\partial \mathcal{M}}=\frac{2}{G_M \kappa \lambda^2}\,\oint\limits_0^{\ell}\dd x^0 \sqrt{\xi} \big(XK-\lambda\sqrt{X}\big)\Big|_{\partial\mathcal M}
    \label{eq:tg21}
\end{equation}
where $K$ is the trace of extrinsic curvature of the cutoff surface $\partial\mathcal M$. Here, the integral over the Euclidean cycle was rewritten such that it is manifestly finite in the TG limit, which requires using the boundary condition $\xi = 1 + \mathcal O(r_c^{-1})$ concurrent with the solutions \eqref{eq:tg18}, and the definition of the canonical ensemble \eqref{eq:tg19}. The inclusion of the boundary action \eqref{eq:tg21} ensures a well-defined Dirichlet problem at the cutoff surface, a well-defined variational principle as the cutoff is removed (\textit{i.e.}, $r_c\to\infty$), and a finite on-shell action.

In both the bulk and the boundary actions only the TG combination $1/\kappa $ of the coupling constants appears as a common overall factor, multiplied by $1/(G_M \lambda^2)$. This allows us to draw two important conclusions: 
\begin{enumerate}
    \item As long as $\lambda $ remains finite when $\epsilon$ goes to zero, the TG limit $\epsilon\to0$ yields a finite prefactor in front of the action.
    \item As long as $\kappa$ is sufficiently small, we can expect a well-defined saddle-point approximation to the Euclidean path integral. 
\end{enumerate}

The Carrollian contraction inherent to the TG limit $\epsilon\to0$ has been performed in \cite{Ecker:2023uwm} along the lines of \cite{Hansen:2021fxi}. To display it, we introduce the inverse partners of the einbein variables defined in the metric \eqref{eq:tg16}, 
$\tau _\mu v^\mu =1$, $e_\mu e^\mu =1$, ${\delta^\mu}_\nu = v^\mu \tau _\nu + e^\mu e_\nu$ 
in terms of which the extrinsic curvature of the cutoff surface is expressed as 
\begin{equation}
    K= -v^\mu e^\nu \big(\partial_{\mu}\tau_{\nu}-\partial_\nu \tau_\mu \big)
    =-2v^\mu e^\nu \partial_{[\mu}\tau_{\nu]} ~ .
    \label{eq:tg23}
\end{equation}
To take the TG limit $\epsilon\to 0$, we expand all quantities in powers of $\epsilon$, denoting the leading order terms by the same letters, which yields the so-called ``magnetic limit'' of the 2d dilaton gravity action \eqref{eq:tg20}--\eqref{eq:tg21}, augmented by the appropriate boundary terms~\cite{SM:2024}:
\begin{widetext}
    \begin{equation}
        \begin{split}
            \frac{1}{\kappa}\Gamma_{\textrm{\tiny TG}} = \lim_{\epsilon\to 0}\Big(\frac{1}{\hbar}\Gamma\Big) &= -\frac{1}{\lambda ^2 G_M \kappa }\int _{\mathcal{M}}\dd ^2x \det (\tau ,e) \Big(XR_{\mathrm{C}}
            +\frac{1}{2X}(e^\mu \partial _\mu X)^2+\frac{\lambda ^2}{2}
            +\gamma _1 \mathscr L_v X+\gamma _2e^\mu \mathscr{L}_ve_\mu\Big) \\
            &\quad -\frac{2}{\lambda ^2 G_M \kappa }\oint_{\partial \mathcal{M}}\!\!\!\dd x^\mu\,\tau_\mu \,\big(2X e^\mu v^\nu  \partial _{[\mu }\tau _{\nu ]}-\lambda \sqrt{X}\big) ~ .
        \end{split}
        \label{eq:carr_action}
    \end{equation}
\end{widetext}

The tantum gravity action \eqref{eq:carr_action} is the main result of our derivation. Besides the typical Carroll dilaton gravity terms in the first line \cite{Grumiller:2020elf,Gomis:2020wxp} --- kinetic and potential terms for the dilaton, and the non-minimal coupling term with the Carroll curvature scalar $R_{\mathrm{C}}$ --- this action also involves two constraints, $\mathscr L_v X=0$ and $e^\mu \mathscr{L}_ve_\mu =0$, and associated Lagrange multipliers, $\gamma_1$ and $\gamma_2$. The constraints are crucial for Carroll-invariance of the action \cite{Ecker:2023uwm,Campoleoni:2022ebj} but play no role thermodynamically since they vanish on-shell. Our new result, the boundary term in the second line of Eq.~\eqref{eq:carr_action}, coincides with the Carrollian limit ($c\to0$) of the boundary term in \cite{Grumiller:2007ju}. 

Notably, the prefactor of the action \eqref{eq:carr_action} depends on the product of the two TG coupling constants \eqref{eq:tg1}. Thus, even though we formally sent $\hbar\to\infty$ in the Bronstein cube, there is still a semi-classical limit corresponding to sufficiently small $\kappa$ \footnote{The semi-classical limit of the action \eqref{eq:whatever} exists whenever the product $\kappa G_M$ is small, which leads to two possible limits, either $\kappa\to 0$ while keeping $G_M$ finite, or $G_M\to 0$ while keeping $\kappa$ finite. However, the latter possibility has to be discarded since it is at odds with our hypotheses to keep the Schwarzschild radius and the energy finite.}. 
In that limit, we can use a saddle-point approximation for the partition function. This amounts to evaluating the action on classical solutions of the theory \eqref{eq:carr_action} and summing over all smooth contributions with a given temperature. Analogously to \eqref{eq:tg18}, the full phase space of \eqref{eq:carr_action} can be labeled by a single constant of motion $r_{\textrm{\tiny S}}$ and is given by 
\begin{subequations}
\label{eq:whatever}
\begin{align}
\tau_\mu&=\sqrt{\xi}\,\delta_\mu^0 & e_\mu&=\frac{1}{\sqrt{\xi}}\,\delta_\mu^1 & v^\mu&=\frac{1}{\sqrt{\xi}}\,\delta^\mu_0\\ e^\mu&=\sqrt{\xi}\,\delta^\mu_1 & X&=\frac{\lambda^2}{4}\,r^2 &  R_{\mathrm{C}}&=\frac{2r_{\textrm{\tiny S}}}{r^3} ~ . &&
\end{align}
\end{subequations}
For any finite  $r_{\textrm{\tiny S}}$ these solutions correspond to Carroll-–Schwarzschild black holes \cite{Perez:2021abf,Hansen:2021fxi,deBoer:2023fnj,Ecker:2023uwm} while for  $r_{\textrm{\tiny S}}=0$ they are just flat Carroll geometries. Since the latter will have a vanishing on-shell action we can drop them and just focus on the black hole sector. Plugging the solutions \eqref{eq:whatever} into the action \eqref{eq:carr_action} establishes the saddle-point approximation of the tantum gravity partition function, 
\begin{equation}
    \ln Z_{\textrm{\tiny TG}} \approx -\frac{1}{\kappa}\,\Gamma_{\textrm{\tiny TG}}\Big|_{\textrm{on-shell}} = -\frac{\beta r_{\textrm{\tiny S}}}{4G_M}=-\beta F ~ .
    \label{eq:angelinajolie}
\end{equation}
With the temperature given by  
$T=\kappa/(4\pi r_{\textrm{\tiny S}})$ \footnote{%
The result for temperature can be derived from the absence of conical defects in the underlying Carrollian manifold; alternatively, the same result follows from a Carrollian version of surface gravity \cite{Ecker:2023uwm}.
}, standard thermodynamic relations, $S=-\partial F/\partial T$ and $E=F+TS$, yield finite results for energy $E=r_{\textrm{\tiny S}}/(2G_M)$ and entropy $S=\pi r_{\textrm{\tiny S}}^2/(\kappa G_M)$. Finally, expressing temperature as a function of energy recovers precisely the Hawking temperature \eqref{eq:tg10}, showing the internal consistency of the tantum gravity limit.

We conclude with a critical assessment of our proposal and derivation, raising first technical and then conceptual points.

We glossed over some technical aspects that already arise for Schwarzschild black hole thermodynamics, namely its negative specific heat and its formally ill-defined canonical ensemble. For the Schwarzschild black hole, this problem was solved by York \cite{York:1986it}, who considered the black hole inside a cavity that couples it to a thermal reservoir, with boundary conditions fixed at the wall of the cavity. Gibbons and Perry showed that the same procedure works in a specific 2d dilaton gravity model \cite{Gibbons:1992rh}, which was later generalized to generic 2d dilaton gravity \cite{Grumiller:2007ju}. While the same issues arise for our tantum gravity example, the Carroll--Schwarzschild black hole, we expect that the same resolution will work as well, \textit{i.e.},  putting it into a cavity with suitable boundary conditions. Alternatively, one could add a negative cosmological constant, which also acts effectively as a cavity \cite{Hawking:1982dh}. It should be worthwhile to work this out in detail for various tantum gravity examples.

Starting from Eq.~\eqref{eq:tg17} we worked in 1+1 dimensions, exploiting spherical reduction of general relativity. The main drawback of such an approach is that it does not generalize straightforwardly to rotating black holes, such as Kerr. Given the difficulties with the Carrollian limit of the Kerr black hole (see the discussion in Section 8 of  \cite{Ecker:2023uwm}), it is an open question how to construct the tantum gravity limit such that rotating black holes are included in the limiting theory. By contrast, a generalization to charged non-rotating black holes, such as Reissner--Nordstr\"om, is straightforward. The main change compared to the results above is that the function $\xi(r)$ in the metric \eqref{eq:tg18} is replaced by $\xi=1-r_S/r+r_q^2/r^2$, where $r_q$ is a new parameter related to the charge, see \cite{Nayak:2018qej,deBoer:2023fnj,Ecker:2023uwm} for more details.

The main advantage of our approach is that it straightforwardly generalizes to arbitrary 2d black hole models, including the tantum gravity limit of Jackiw--Teitelboim (JT) \cite{Teitelboim:1983ux,Jackiw:1984je} or Witten black holes \cite{Elitzur:1991cb,Mandal:1991tz,Witten:1991yr}. It is worthwhile to study several aspects of the Sachdev--Ye--Kitaev/JT-correspondence \cite{Kitaev:15ur,Sachdev:1992fk,Sachdev:2010um,Jensen:2016pah,Maldacena:2016upp,Cotler:2016fpe,Engelsoy:2016xyb,Mertens:2018fds,Sarosi:2017ykf,Gu:2019jub,Saad:2019lba} in the tantum gravity limit, starting with the Schwarzian boundary action, relations to matrix models, and aspects of the information paradox (see, \textit{e.g.}, \cite{Penington:2019npb,Almheiri:2019psf,Almheiri:2019hni,Almheiri:2019yqk,Almheiri:2019qdq,Penington:2019kki}). The bulk-plus-boundary action \eqref{eq:carr_action} will play a key role in any such discussion. Additionally, it will be good to understand what the tantum gravity limit implies on the field theory side in a holographic correspondence and how it affects the universal Schwarzian sector~\footnote{%
This universal sector has a holographic interpretation in terms of the strongly coupled gravitational dynamics in the throat of near-extremal black holes. For non-extremal black holes there may be a similar story of universal near-horizon dynamics, based on the twisted warped Schwarzian boundary action discovered in \cite{Afshar:2019tvp} and applied in \cite{Afshar:2019axx,Godet:2020xpk}.
} in two-dimensional conformal field theories \cite{Ghosh:2019rcj}. Lower-dimensional tantum gravity models could play a decisive role in the conceptual understanding of the tantum gravity limit \eqref{eq:whynot} in a holographic context. 

In the saddle-point approximation $\kappa\to0$, TG reduces to (magnetic) Carroll gravity so that one can think of the latter as a ``classical'' version of TG. We left open an investigation of the dual limit TG$^\ast$, which we expect to have an action of Galilean type \cite{Hartong:2022lsy} as its saddle-point approximation. Thus, one could refer to tantum gravity as ``Carrollian tantum gravity'' and to its dual as ``Galilean tantum gravity.'' However, this nomenclature erroneously suggests the existence of some Lorentzian tantum gravity theory that would smoothly interpolate between these two limits. From the Bronstein cube, it is clear that this cannot be true since these limiting cases are antipodal, so the only smooth way to connect TG and TG$^\ast$ goes through quantum gravity.

Furthermore, our results seem in tension with the claim of \cite{deBoer:2023fnj} that thermal ``partition functions of Carroll systems are ill-defined and do not lead to sensible thermodynamics.'' This conclusion was reached because they kept $\hbar$ finite in the Carrollian limit. However, as we have derived inductively in our Letter, we can insist on sensible thermodynamics if we scale $\hbar\to\infty$ while sending $c,G_N\to 0$. 

Finally, we hope that our tantum gravity proposal will contribute to an overdue development of Carroll thermodynamics from first principles and to new insights into the elusive theory of quantum gravity. 


\begin{acknowledgments}
\section*{Acknowledgments} 
We thank Arjun Bagchi, Jan de Boer, Jelle Hartong, Emil Have, Niels Obers, Alfredo P\'erez, Stefan Prohazka, Romain Ruzziconi, Ricardo Troncoso and Stefan Vandoren for useful discussions.

This work was supported by the Austrian Science Fund (FWF), projects P~32581, P~33789, and P~36619.    
\end{acknowledgments}


\renewcommand{\em}{\it}

\providecommand{\href}[2]{#2}\begingroup\raggedright\endgroup

\end{document}